\documentclass[12pt]{iopart}

\usepackage{graphicx}
\begin{document}

\title[LArSoft]{LArSoft: A Software Package for Liquid Argon Time Projection Drift Chambers}

\author{Eric Church}
\address{Yale University, PO Box 500, MS309, Fermi National Accelerator Lab, Batavia, IL, USA, 60510-5011}
\ead{echurch@fnal.gov}

\begin{abstract}
We describe in these GLA2011 proceedings the software package LArSoft, a toolkit to perform simulation, analysis and reconstruction with the Liquid Argon (LAr) Time Projection Chambers (TPCs) within the US program of proposed detectors. We demonstrate that LArSoft is a fast-maturing, sophisticated package which has taken on important analyses already, and which stands ready to be adopted by as many as five Liquid Argon Time Projection Chambers.

\end{abstract}

\maketitle

\section{Introduction }
The imposing task of building a software package to perform Simulation, Reconstruction and Analysis on a many-tens-of thousand channel TPC is organized under one banner in the US LAr program. That program is LArSoft. LArSoft is engineered to be agnostic to precisely which detector runs in its framework. The detectors enivisioned at this time to avail themselves of LArSoft are ArgoNeuT, MicroBooNE, and the future LBNE LAr detector, as well as detectors in various testbeam and non-beam Research and Development efforts. One imagines that if the proposed 1 ktonne, pre-LBNE prototype, LAr1, becomes a reality its collaborators will use LArSoft. The engineering prototype Bo also proposes to use the tools of LArSoft. LArSoft is headquartered at http://cdcvs.fnal.gov/redmine/projects/larsoftsvn/.

\section{Fermilab Support}
LArSoft rests on the Fermilab-supported Analysis Reconstruction Tools (ART) framework, itself an outgrowth of the CMSSoftWare framework (CMSSW). ART is a straightfoward framework that allows access to an event record and to build  events up into collections of events. Simulation, reconstruction and analysis use a handful of mechanisms to get at these events. There are enormous gains to be had by building one's analysis tools on a supported framework. ART is used by other Fermilab Intensity Frontier experiments Mu2e and NOvA and certainly will be used by future Intensity Frontier programs, as well. The ART team at Fermilab makes new features and bug-fixes available, and for the most part LArSoft carries on its business, transparent to such changes. ART is a complicated state machine attended to by Fermilab's Computing Division who pay attention to the low level details, freeing LArSoft workers to the detector and physics matters at hand.
Similarly, Fermilab tends to the externals against which LArSoft is built. Thus, externals like ROOT~\cite{ROOT} and GEANT4~\cite{GEANT4}, etcetera, packages LArSoft uses at deep levels, are kept up-to-date and made available in what are called relocatable UPSes. LArSoft conveners therefore do not need to keep up with every new ROOT release, downloading and building these packages and swapping them in for the old versions. LArSoft conveners merely point at the new versions of ART, ROOT, and other externals, and the LArSoft code is automatically built in a simple cutover against the latest, greatest tools. 

The build system, SoftRelTools (SRT), is also supported by Fermilab. Fermilab's condor grid is used for batch computing by LArSoft. Finally, the LArSoft code repository, svn, works seamlessly with Fermilab's project management tool of choice, redmine.

\section{Run Mechanics}

\subsection{Geometry}
For the detector under study two things must be specified. One is the appropriate Geant4 gdml geometry file. The second is the detector electronics package. This includes data format particular to the detector, but also response files which allow raw time signals from Collection and Induction planes from a given detector to be pedestal-subtracted and deconvolved and converted into unipolar signals. Those signals then are subjected to hit-finding algorithms.

\subsection{Events} 
In the case of ArgoNeuT which has been exposed to the Fermilab Main Injector beam, raw data files can be specified as the input. Monte Carlo events, meanwhile, are created in the detector under study by one of four mechanisms. Particles whose responses we want to track and study in the simulation come into being via 

\begin{itemize}
\item the neutrino interaction generators GENIE~\cite{GENIE} (code supported by its authors) or NuANCE~\cite{NuANCE} (now unsupported), or by 
\item the application of the cosmic ray event generator CRY~\cite{CRY}, or
\item single particle mode generation, or 
\item are read in from Files.
\end{itemize}

GENIE/NuANCE produce neutrino secondaries according to flux files appropriate to the detector under study. CRY secondaries rain down on a detector according to the cosmic ray model given by~\cite{CRY.model}. All particle kinematics and locations of origin in single particle mode may be specified. In file mode, those parameters are read in from the file.

\section{LArSoft Processor Flow and Components}

\begin{figure}[h]
\centering
\caption{This flow chart shows the objects created along the reconstruction chain in LArSoft. The optical simulation/reconstruction chain is not included here.}
\includegraphics[width=4.0in]{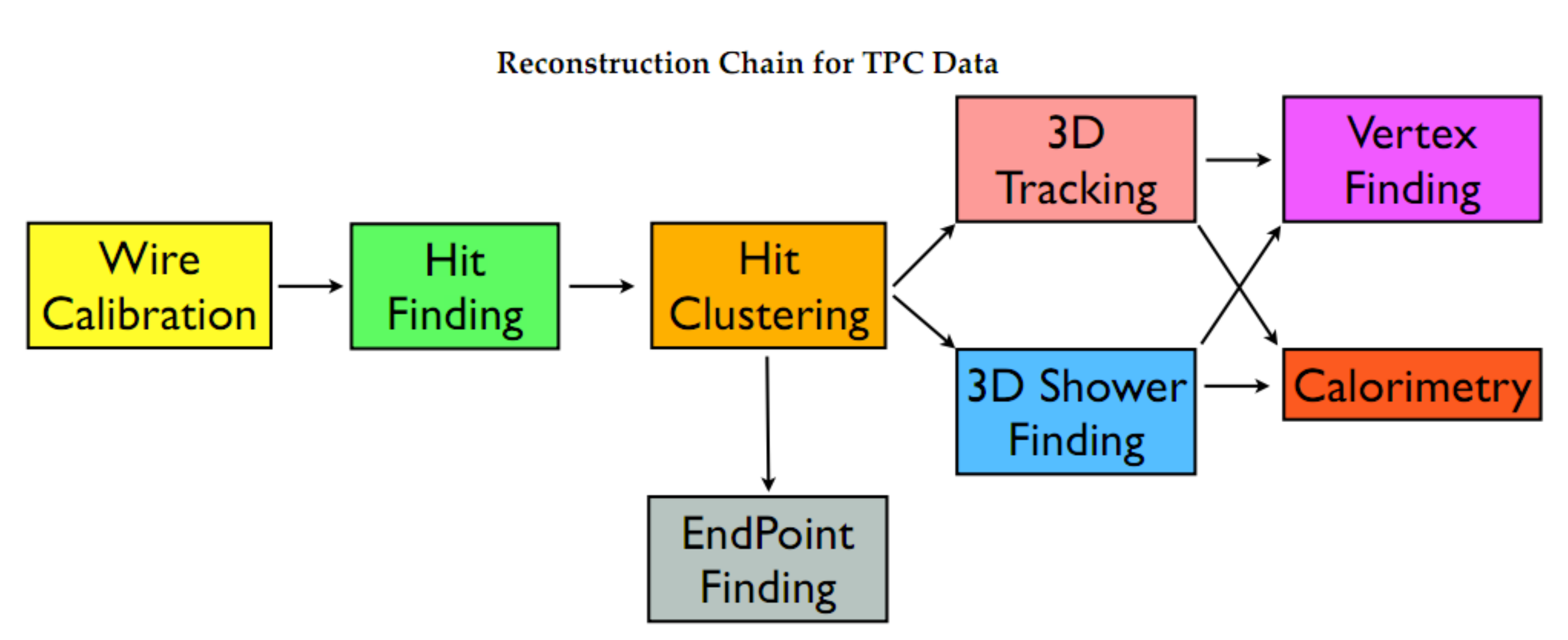}
\label{flow}
\end{figure}

\begin{figure}[h]
\centering
\caption{This event display shows some of the LArSoft reconstruction objects in a typical $\pi^0$ decay in an ArgoNeuT Monte Carlo event. Upper (middle) panel shows Collection (Induction) plane time versus wire activity. Hits on individual wire planes are aggregated into clusters. These 2D clusters are made into 3D showers. The bottom panel shows the black uncalibrated induction signal on a particular wire, the blue is the deconvolved calibrated pulse as obtained in the Wire Calibration module, and the red box is a hit as determined from the Hit Finder.}
\includegraphics[width=4.0in]{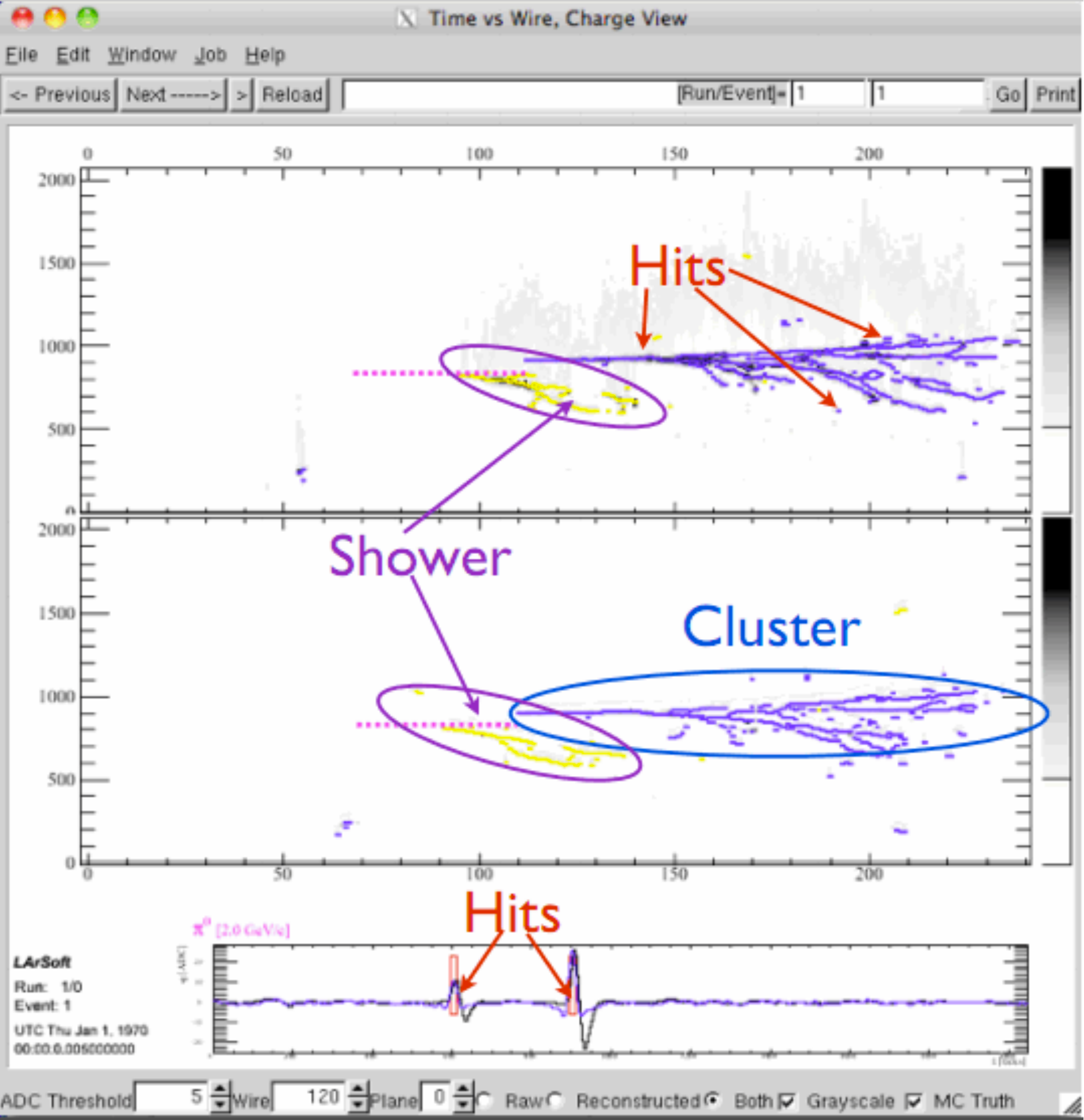}
\label{cheat}
\end{figure}

Figure~\ref{flow} shows the LArSoft flow chart. Processing follows through the boxes according to the arrows there. Figure~\ref{cheat} shows an event display for some of the objects that those boxes create. Next, we elucidate some of the processes mentioned in the boxes of figure~\ref{flow}.

\hspace*{2cm}

\hspace*{2cm}

\subsection{Calibrated Data}
This module reads up the raw wire data, as previously mentioned and outputs pulses for the Hit Finder. 

\subsection{Hit Finder}
The hit finder looks for pulses and applies N-Guassian fits to the pulses, with $N\ge1$ as the algorithm warrants. Hits are counted, and charge is accumulated under the peaks and peak positions are stored away for future use.

\subsection{Cluster Finder}
The cluster finder puts hits together according to their proximity to one another, per the DBSCAN~\cite{DBSCAN} algorithm.

\subsection{2D Line Finder}
This module applies a Hough transform~\cite{Hough} to find all line segments within a cluster. Lines may later be merged if their slopes and endpoints are similar.

\subsection{Line Endpoint Finder}
In this module endpoints are found for the line segments of the previous module. They may be use as seeds for real 3 dimensional vertices in subsequent processing.

\subsection{3D Track Finder}
This module is the first point at which two dimensional, single plane information is merged to produce three dimensional objects. 2D tracks in this module are abstracted to 3D tracks via one or more algorithms. Those algorithms include projecting up from the overlapped hits on the 2D Lines to the real 3D track that produced those hits, as well as a Kalman track fitter which finds the best non-parametrized track that can be run through space points which are not necessarily forced to sit on straight lines.

\subsection{3D Shower Finder}
The 3D shower finder defines a vertex of a given cluster, and projects two of the clusters in any two planes into a 3D shower starting at that vertex. The best pairing of the three possibilities is kept.

\subsection{3D Vertex Finder}
This module takes the tracks from the Kalman filter and projects them back to their distances of closest approach and asks if a vertex can be formed to within the pointing errors.

\subsection{Event Finder}
The event finder gathers up all vertices and its daughter objects and asks which can be reasonably associated with one event. One can imagine this has particular application in disentangling overlaid cosmic events in MicroBooNE or piled up neutrino events in the near detector for LBNE.

\subsection{Tracking Photons}
A full optical photon simulation exists in LArSoft. In LAr tens of thousands of 128~nm photons are produced per MeV of deposited energy. MicroBooNE will employ a wavelength shifting technology along with photomultiplier tubes (PMTs) to capture that light to give a time=0 to its beam events. LBNE is still discussing its light collection options.

In LArSoft optical photons are created at a grid of points in the simulated MicroBooNE TPC geometry. The response at each of the 30 PMTs due to a light source at each point in that grid is measured from one large simulation at each point, and a look-up table is created. Rayleigh scattering and other material considerations are taken into account. The look-up table is used in subsequent runs to study light collection for scenarios of bad PMTs, etcetera.

\section{Studies in ArgoNeuT}

ArgoNeuT, with its $1.35e20$ protons on target with neutrino and antineutrino beams is at the forefront of LArSoft development.

\subsection {Electron Lifetime}
Electron lifetime studies -- the attenuation with distance due to electron recombination -- are performed by fitting a Landau distribution to charge collected in bins of drift time for through-going muons. A lifetime $\tau$ is then fitted to the most probable charge value versus drift time.

\subsection{Particle Identification}
Particle Identification is performed once 3D Showers are available by plotting dE/dx vs residual range. ArgoNeuT has shown~\cite{josh} how within an individual track the individual hits of likely muon and proton tracks fall consistently within the MC expected bands for those hypotheses.

\section{An LBNE/LAr40 Study}
LAr40 is the 40 ktonne LAr detector proposed for the 800 foot level at the Deep Underground Science Lab (DUSEL) in South Dakota. In the talk~\cite{mytalk} we show how a nucleon decay study may be undertaken in LArSoft with GEANT4. Event biasing tactics are described which allow producing rare decays. This is relevant because cosmic muon-induced spallation will lead to the irreducible background for proton decay golden mode in LAr, namely $p\rightarrow K^+\overline{\nu}$. We re-toss secondaries, bias the cross sections, turn on charge exchange for K0s, and make strict choices about which particles on the GEANT4 stack to track and which to kill. We show that the rate of cosmic muons entering LAr40, $\approx 250$ muons/second, is consistent with that quoted in reference \cite{bueno}.

\section{Conclusions}
LArSoft is a fast-maturing software analysis, simulation and reconstruction toolkit which is built upon the ART framework and extensive Fermilab support, generally. The first dissertation\cite{josh} from ArgoNeuT, a measurement of the charged current inclusive anti-neutrino cross-section in the Main Injector beam at Fermilab, is recently out. That analysis is performed using an automated selection, reconstruction and simulation all within LArSoft.

It is important to demonstrate LArSoft's usefulness and efficacy at performing analyses next on reconstructed, simulated MicroBooNE events, with its $\approx 20\times$ increase in number of channels,  and $\approx 4\times$ increase in event time samples and $\approx 700\times$ active volume increase. Most of the reconstruction chain is in place and development continues at a fast pace. There is much at stake in the effort to provide productive and efficient software tools for the US LAr TPC program.

\end{document}